\documentclass[twocolumn,showpacs,amsmath,amssymb,aps,prl]{revtex4}
\def\be{\begin{equation}}
\def\ee{\end{equation}}

\def\bea{\begin{eqnarray}}
\def\eea{\end{eqnarray}}

\def\e{\epsilon}

\begin{document}
%\begin{fmffile}{pics}
\pagestyle{empty}
\preprint{MANUSCRIPT}

\title{Time-dependence of correlation functions following a quantum quench}

\author{Pasquale Calabrese}
\affiliation{Institute for Theoretical Physics, University of Amsterdam,
	Valckenierstraat 65, 1018 XE Amsterdam, The Netherlands}
\author{John Cardy}
\affiliation{ Oxford University, Rudolf Peierls Centre for
Theoretical Physics, 1 Keble Road, Oxford, OX1 3NP, United
Kingdom} \affiliation{All Souls College, Oxford, United Kingdom}

\date{\today}% It is always \today, today,
             %  but any date may be explicitly specified

\begin{abstract}
We show that the time-dependence of correlation functions in an
extended quantum system in $d$ dimensions, which is prepared in
the ground state of some hamiltonian and then evolves without
dissipation according to some other hamiltonian, may be extracted
using methods of boundary critical phenomena in $d+1$ dimensions.
For $d=1$ particularly powerful results are available using
conformal field theory. These are checked against those available
from solvable models. They may be explained in terms of a picture,
valid more generally, whereby quasiparticles, entangled over
regions of the order of the correlation length in the initial
state, then propagate classically through the system.

\end{abstract}

\pacs{73.43.Nq, 11.25.Hf. 64.60.Ht} 
                             % PACS, the Physics and Astronomy
                             % Classification Scheme.
%\keywords{Suggested keywords}%Use showkeys class option if keyword
                              %display desired
\maketitle

Suppose that an extended quantum system in $d$ dimensions (for
example a quantum spin system), is prepared at time $t=0$ in a
pure state $|\psi_0\rangle$ which is the ground state of some
hamiltonian $H_0$ (or, more generally, in a thermal state at a
temperature less than the gap $m_0$ to the first excited state.)
For times $t>0$ the system evolves \em unitarily \em according to
the dynamics given by a \em different \em hamiltonian $H$, which
may be related to $H_0$ by varying a parameter such as an external
field. This variation, or quench, is supposed to be carried out
over a time scale much less than $m_0^{-1}$. How do the
correlation functions, expectation values of products of local
observables, then evolve? The answer to this question would appear
to depend in detail on the system under consideration. It was
first addressed in the context of the quantum Ising-XY model in
Refs.~\cite{bm1} (see also \cite{ir-00}). Until recently it
was, however, largely an academic question, because the time
scales over which most condensed matter systems can evolve
coherently without coupling to the local environment are far too
short, and the effects of dissipation and noise are inescapable.
However, with the development of experimental tools for studying
the behavior of optical lattices of ultracold atoms, and quantum
phase transitions in these systems \cite{uc}, there has been
renewed interest in this theoretical problem (see, for example,
\cite{sps-04,cl-05}.)

In this Letter we study such problems in general and argue that,
if $H$ is at or close to a quantum critical point (while $H_0$ is
not), there is a large degree of universality in the behavior at
sufficiently large distances and late times, despite the fact that
correlations typically fall off exponentially, rather than the
power laws characteristic of the ground state near a quantum
critical point. Our arguments are based on the path integral
approach and the well-known mapping of the quantum problem to a
classical one in $d+1$ dimensions. The initial state plays the
role of a boundary condition, and we are able to then use the
renormalisation group (RG) theory of boundary critical behavior
(see, e.g., \cite{dd}). From this point of view,
particularly powerful analytic results are available for $d=1$ and
when the quantum critical point has dynamic exponent $z=1$ (or,
equivalently, a linear quasiparticle dispersion relation
$\omega=v|k|$) because then the $1+1$-dimensional problem is
described asymptotically by a boundary conformal field theory
(BCFT)\cite{cardy-84,cardy-05}. Some of these methods have
recently been applied \cite{cc-05} to studying the time evolution
of the entanglement entropy, but, as we shall argue, they are more
generally applicable. Further details of these calculations will
appear elsewhere \cite{cc-06}.

The results we find from CFT suggest a rather simple picture which
is, however, more generally applicable: the state
$|\psi_0\rangle$, which has an (extensively) high energy compared
with that of the ground state of $H$, acts as a source for
quasiparticle excitations. Those quasiparticles originating from
closely separated points (roughly within the correlation length
$\xi_0$ of the ground state of $H_0$) are quantum entangled.
However, once they are emitted they behave semi-classically,
travelling at speed $v$. They have two distinct effects. Firstly,
incoherent quasiparticles arriving a given point $\bf r$ from
well-separated sources cause relaxation of (most) local
observables at $\bf r$ towards their ground state expectation
values. (An exception is the local energy density which of course
is conserved.) In the CFT case this relaxation is exponential
$\sim\exp(-\pi xvt/2\tau_0)$. Here $\tau_0\sim m_0^{-1}$ is
non-universal, but $x$, the bulk \em scaling dimension \em of the
particular observable, is related to the critical exponents of the
quantum phase transition, and is universal. Hence \em ratios \em
of decay constants for different observables should be universal.
Secondly, entangled quasiparticles arriving at the same time $t$
at points with separation $|{\bf r}|\gg\xi_0$ induce (quantum)
correlations between local observables. In the case where they
travel at a unique speed $v$, therefore, there is a sharp `light
cone' effect: the connected correlations do not change
significantly from their initial values until time $t\sim |{\bf
r}|/2v$. In the CFT case this light-cone effect is rounded off in
a (calculable) manner over the region $t-|{\bf r}|/2v\sim\tau_0$,
since quasiparticles remain entangled over this distance scale.
After this they rapidly saturate to \em time-independent \em
values. For large separations (but still $\ll 2vt$), these decay
\em exponentially \em $\sim\exp(-\pi x|{\bf r}| /2v\tau_0)$. Thus,
while the generic one-point functions relax to their ground-state
values, the correlation functions do not, because, at quantum
criticality, these would have a power law dependence. Of course,
this is to be expected since the mean energy is much higher than
that of the ground state, and it does not relax.

These results rely on the technical assumption that the leading
asymptotic behavior given by CFT, which applies to the euclidean
region (large imaginary times), may simply be analytically
continued to find the behavior at large real time. While such
procedures have been shown to give the correct behavior for the
time-dependent correlations in equilibrium, it is important to
check them in specific solvable cases. We have done this for the
case of a lattice free boson (coupled harmonic oscillators), and
for the Ising-XY chain, which can be transformed into a free
fermion problem. These models, which are exactly solvable on the
lattice, confirm our general results and also show how the
semi-classical picture discussed above is modified for a more
general quasiparticle dispersion relation $\omega=\Omega_k$,
taking into account both the effects of the lattice and of a
finite gap. The results are consistent with this picture as long
as the quasiparticles are assumed to propagate at the group
velocity $\Omega'_k$ appropriate to their wave number $k$. In this
case the light-cone effect first occurs at time $t\sim |{\bf
r}|/2v_m$, where $v_m$ is the maximum group velocity. (If this
occurs at a non-zero wave number, it gives rise to spatial
oscillations in the correlation function.) However, because there
are also quasiparticles moving at speeds less than $v_m$, the
approach to the asymptotic behavior at late times is less abrupt.
In fact, for a lattice dispersion relation where $\Omega'_k$
vanishes at the zone boundary, the approach to the limit is slow,
as an inverse power of $t$. A similar result applies to the
1-point functions. This is consistent with the exact results found
in \cite{bm1}.

We first discuss some general features of the path integral
approach and its relation to boundary critical behavior. The
expectation value of any product of local operators is given by
\begin{equation}
\langle {\cal O}(t,\{ {\bf r}_i\})\rangle= Z^{-1} \langle \psi_0 |
e^{i H t-\e H} {\cal O}(\{ {\bf r}_i\}) e^{-i H t-\e H}| \psi_0
\rangle \label{Oexp}
\end{equation}
where we have included damping factors $e^{-\e H}$ in such a way
as to make the path integral representation of the expectation
value absolutely convergent. The normalization factor is
$Z=\langle\psi_0|e^{-2\e H}|\psi_0\rangle$. Eq.~(\ref{Oexp}) may
be represented by a path integral in imaginary time $\tau$
\begin{equation}
\label{pi} \frac1Z\int[d\phi({\bf r},\tau)]{\cal O}(\{{\bf
r}_i\},0) \,e^{-S[\phi]} \langle\psi_0|\phi({\bf r},\tau_2)\rangle
\langle\phi({\bf r},\tau_1)|\psi_0\rangle
\end{equation}
over a complete set of fields $\phi({\bf r},\tau)$ (or, in a spin
system, a coherent state representation), with
$S=\int_{\tau_1}^{\tau_2}Ld\tau$, analytically continued to
$\tau_1=-\e-it$ and $\tau_2=\e-it$. Here $L$ is the (euclidean)
lagrangian corresponding to the dynamics of $H$. We will consider
the equivalent slab geometry between $\tau=0$ and $\tau=2\e$, with
$\cal O$ inserted at $\tau=\e+it$.

Eq.~(\ref{pi}) has the form of the equilibrium expectation value
in a $d+1$-dimensional slab geometry with particular boundary
conditions. We wish to study this in the limit when $t$ and the
separations $|{\bf r}_i-{\bf r}_j|$ are much larger than the
microscopic length and time scales, when RG theory can be applied.
If $H$ is at or close to a quantum critical point, the bulk
properties of the critical theory are described by a bulk RG fixed
point (or some relevant perturbation thereof). In that case, the
boundary conditions flow to one of a number of possible \em
boundary \em fixed points \cite{dd}. Thus, for the purpose of
extracting the asymptotic behavior, we may replace
$|\psi_0\rangle$ by the appropriate RG-invariant boundary state
$|\psi_0^*\rangle$ to which it flows. The difference may be taken
into account, to leading order, by assuming that the RG-invariant
boundary conditions are not imposed at $\tau=0$ and $\tau=2\e$ but
at $\tau=-\tau_0$ and $\tau=2\e+\tau_0$. In the language of
boundary critical behavior, $\tau_0$ is called the extrapolation
length \cite{dd}. It characterizes the RG distance of the actual
boundary state from the RG-invariant one. It is always necessary
because scale-invariant boundary states are not in fact
normalizable\cite{cardy-05}. It is expected to be of the order of
the typical time-scale of the dynamics near the ground state of
$H_0$, that is the inverse gap $m_0^{-1}$. The effect of
introducing $\tau_0$ is simply to replace $\e$ by $\e+\tau_0$. The
limit $\e\to0+$ can now safely be taken, so the width of the slab
is then taken to be $2\tau_0$.

\noindent\em $d=1$ and CFT\em. The above was completely general,
but we now consider the case when $H$ is at a quantum critical
point whose long-distance behavior is given by a 1+1-dimensional
CFT, with dispersion relation $\omega=v|k|$. We set $v=1$ in the
following. RG-invariant boundary conditions then correspond to
conformally invariant boundary states. Under these circumstances
the correlation functions of local operators in the slab geometry
(whose points are labelled by a complex number $w$ with $0<{\rm
Im} w<2\tau_0$) are related to those in a half-space ${\rm Im}
z>0$ by the conformal mapping $w=(2\tau_0/\pi)\log z$. In the case
where $\cal O$ is a product of local \em primary \em scalar
operators $\Phi_i(w_i)$ (which covers most of the physically
relevant cases, for important exceptions see below) we have
\begin{equation}
\label{stripUHP} \langle\prod_i\Phi_i(w_i)\rangle_{\rm strip}
=\prod_i|w'(z_i)|^{-x_i} \langle\prod_i\Phi_i(z_i)\rangle_{\rm
UHP}
\end{equation}
where $x_i$ is the bulk scaling dimension of $\Phi_i$. Note that
the expectation values of the $\Phi_i$ in the ground state of $H$
are supposed to have been subtracted off. We now discuss some
special cases of (\ref{stripUHP}).

\noindent\em One-point functions\em. In this case scale invariance
implies that $\langle\Phi(z)\rangle_{\rm UHP}=A^\Phi_b(2{\rm Im}
z)^{-x}$, where $A^\Phi_b$ is an amplitude depending on the
selected boundary condition. (It may vanish if $\Phi$ corresponds
to an operator whose expectation value in $|\psi_0\rangle$
vanishes.) This gives
\begin{equation}
\langle\Phi(w)\rangle_{\rm strip}= A^\Phi_b
\left[\frac{\pi}{4\tau_0} \frac{1}{\sin(\pi
\tau/(2\tau_0))}\right]^{x}\,.
\end{equation}
Continuing to $\tau=\tau_0+ it$ we then find
\begin{equation}
\label{decay} \langle\Phi(t)\rangle= A^\Phi_b
\left[\frac{\pi}{4\tau_0} \frac{1}{\cosh(\pi
t/(2\tau_0))}\right]^{x}\,,
\end{equation}
which exhibits exponential decay for $t\gg\tau_0$, with a decay
time related to the critical exponent $x$, as described in the
introduction.

An important exception to this law is the local energy density (or
any piece thereof). This corresponds to the $tt$ component of the
energy-momentum tensor $T_{\mu\nu}$. In CFT this is not a primary
operator. Indeed, if it is normalised so that $\langle
T_{\mu\nu}\rangle_{\rm UHP}=0$, in the strip \cite{cardybn-86}
$\langle T_{tt}({\bf r},\tau)\rangle=\pi c/24(2\tau_0)^2$ (where $c$ is
the central charge  of the CFT) so that it does not decay in time.
Of course this is to be expected since the dynamics conserves
energy. A similar feature is expected to hold for other local
densities corresponding to globally conserved quantities which
commute with $H$, for example the total spin.

\noindent\em Two-point functions\em. The 2-point function in the
half-plane takes the general form \cite{cardy-84}
\begin{equation}
\langle\Phi(z_1) \Phi(z_2) \rangle_{\rm UHP}=\left
(\frac{z_{1\bar2}z_{2\bar1}}{z_{12}z_{\bar1\bar2}z_{1\bar1}z_{2\bar2}}\right)^{x}
F(\eta)\,, \label{2ptgen}
\end{equation}
where $z_{ij}=z_i-z_j$, $z_{\bar i}$ is the image of $z_i$ in the
real axis, and $\eta\equiv
z_{1\bar1}z_{2\bar2}/z_{1\bar2}z_{2\bar1}$. The function $F$ is
universal but depends in detail on the particular BCFT. Using
(\ref{stripUHP}), analytically continuing, and assuming
$r,t\gg\tau_0$ we find
\begin{equation}
\langle\Phi(0,t) \Phi(r,t) \rangle=
(\pi/2\tau_0)^{2x}\left(\frac{e^{\pi r/2\tau_0} +e^{\pi t/\tau_0}}
{e^{\pi r/2\tau_0}\cdot e^{\pi t/\tau_0}}\right)^{x}\,F(\eta)\,,
\end{equation}
where now
\begin{equation}
\eta\sim\frac{e^{\pi t/\tau_0}}{e^{\pi r/2\tau_0}+e^{\pi
t/\tau_0}}\,.
\end{equation}
Thus for $r-2t\gg\tau_0$, $\eta\sim e^{\pi(t-r/2)/\tau_0}\ll1$,
while for $2t-r\gg\tau_0$, $\eta\sim1$. In both of these limits
the behavior of $F$ is determined by the short-distance
expansion. The first limit, when $t<r/2$, corresponds in the
half-plane to both points being close to the boundary, when we can
use the bulk-boundary expansion to argue that $F(\eta)\sim
(A^\Phi_b)^2\eta^{x_b}$, where $x_b$ is the boundary scaling
dimension of the leading boundary operator to which $\Phi$
couples. This gives, for $t<r/2$,
\begin{equation}
\label{lessthan} \langle\Phi(r,t)\Phi(0,t)\rangle\sim (A^\Phi_b)^2
e^{-\pi xt/\tau_0}\times e^{-\pi x_b(r/2-t)/\tau_0}\,.
\end{equation}
Note that if $\langle\Phi\rangle\not=0$, $x_b=0$ and
(\ref{lessthan}) is just $\langle\Phi\rangle^2$. In that case the
leading behavior of the \em connected \em 2-point function is
given by subleading terms either in $F$ or in the bulk-boundary
short-distance expansion. The opposite limit, when $t>r/2$,
corresponds to the two points both being far from the boundary, in
which case the bulk behaviour dominates and $F\to1$. This gives
\begin{equation}
\label{greaterthan} \langle\Phi(r,t)\Phi(0,t)\rangle\sim e^{-\pi
xr/2\tau_0}\,,
\end{equation}
that is the correlations saturate (exponentially fast) for
$t>r/2$. Note, however, that the precise details of the cross-over
behavior for $|t-r/2|\sim\tau_0$ are dependent on the form of $F$
for $0<\eta<1$.

\noindent\em Other CFT results\em. Many other general results may
be found within the CFT formalism. For example, the 2-time
correlation function
\begin{equation}
\langle\Phi(r,t)\Phi(0,s)\rangle \sim
\begin{cases}
e^{-\pi x(t+s)/4\tau_0}
\; &{\rm for}\, r>t+s\,,\\
e^{-\pi xr/4\tau_0}\; &{\rm for}\, t-s<r<t+s\,,\\
e^{-\pi|t-s|/4\tau_0}\; &{\rm for}\, r<|t-s|\,.
\end{cases}
\end{equation}
where it is assumed that $t$, $s$, $|t-s|$ and $r$ are all
$\gg\tau_0$. In the case where $\langle\Phi\rangle=0$, the first
case gains an additional factor $e^{-\pi x_b(t+s-r)/4\tau_0}$.
Note that the autocorrelation function with $r=0$ depends only on
the time difference and does not exhibit aging in this regime.
Another example is the 1-point function in a semi-infinite chain,
for which we find
\begin{equation}
\langle\Phi(r,t)\rangle\sim
\begin{cases}
e^{-\pi xt/2\tau_0}\quad &{\rm for}\, t<r\,,\\
e^{-\pi xr/2\tau_0}\quad &{\rm for}\, t>r\,.
\end{cases}
\end{equation}

\noindent\em The gaussian chain\em. The simplest exactly solvable
model in which to study these effects is a general chain of
coupled harmonic oscillators with a hamiltonian
\begin{equation}
\frac12\sum_r\Big(\pi_r^2+m^2\phi_r^2+\sum_j\omega_j^2(\phi_{r+j}-\phi_r)^2
\Big)
\end{equation}
with the standard commutation relations imposed between the
$\phi_r$ and their canonical conjugates $\pi_r$. The 2-point
function can be found straightforwardly if tediously by
integrating the Heisenberg equations of motion for the Fourier
modes. The result is
\begin{eqnarray}
 \langle\phi_r(t)\phi_0(t)\rangle&-&
\langle\phi_r(0)\phi_0(0)\rangle\label{2ptgaussian}\\
&=& \int_{\rm BZ} e^{ikr}
\frac{(\Omega_{0k}^2-\Omega_k^2)(1-\cos(2\Omega_kt))}
{\Omega_k^2\Omega_{0k}}dk \nonumber
\end{eqnarray}
where $\Omega_{0k}$ and $\Omega_{k}$ are the single-particle
dispersion relations corresponding to $H_0$ and $H$ respectively,
and the integral is over the first zone. The CFT results
correspond to the case with $\Omega_k\sim v|k|$ as $k\to0$, which
gives $\langle\phi_r(t)\phi_0(t)\rangle\sim m_0(t-r/2v)$ for
$m_0(t-r/2v)\gg1$, while it vanishes for $t<r/2v$. Taking $\Phi$
to be the conformal field $e^{iq\phi}$ then reproduces all the CFT
results before, with $\tau_0\sim m_0^{-1}$, $x\propto q^2$ and
$F=1$.

However, the advantage of this simple model is that the effects of
other dispersion relations can be understood. In fact, the
dominant contribution to (\ref{2ptgaussian}) in the limits of
large $t$ and $r$ comes from where, by a stationary phase
argument, the group velocity $v_k\equiv\Omega'_k=r/2t$. This
suggests the picture given in the introduction of the correlations
being due to left- and right-moving particles emitted at $t=0$
from closely separated points, moving with the group velocity. It
also explains for the case of a semi-infinite chain why the
relevant time scale is $r/v$ rather than $r/2v$, since one of the
particles is reflected from the end of the chain. For the energy
density, proportional to $(\phi_{r+1}-\phi_r)^2$, the asymptotic
behavior is a constant, but the approach to this is dominated by
the particles with the smallest group velocity, which, for a
gapless lattice model, come from the zone boundary at $|k|=\pi$.
An explicit calculation\cite{cc-06} gives a correction $\sim
t^{-3/2}\cos(\Omega_\pi t+\pi/4)$. Similarly, for a quench to a
\em gapped \em lattice $H$, with $\Omega_0>0$, the fastest
particles correspond to a non-zero wave number. This gives rise to
spatial oscillations in the correlation function.

\noindent\em The Ising-XY chain\em. As an example consider the CFT
describing the scaling limit of the Ising-XY chain with
hamiltonian
\begin{equation}
\sum_{r=1}^N \Big(\frac{(1+\gamma)}{2} \sigma^x_r \sigma_{r+1}^x
+\frac{(1-\gamma)}{2}\sigma^y_r \sigma_{r+1}^y-
h\sigma_r^z\Big)\,,
\end{equation}
where $\gamma$ is the anisotropy parameter and $h$ is the
external transverse field. It is well known that for any
$\gamma\neq0$ the model undergoes a phase transition at $h=1$ that
is in the universality class of the Ising model (defined by
$\gamma=1$). In that case $F$ is known exactly \cite{cardy-84} to
be $F=(\sqrt{1+\eta^{1/2}}\pm\sqrt{1-\eta^{1/2}})/\sqrt2$, where
the upper and lower signs correspond respectively to fixed and
free boundary conditions on the spins $\sigma_r^x$. These
correspond to quenches to the critical point $h=1$ from $h_0<1$
and $h_0>1$ respectively. Our general results should then apply, for 
example, to the correlations of the order parameter
$\sigma^x$, with scaling dimensions $x=\frac18$ and $x_b=\frac12$
for $h_0>1$ and zero for $h_0<1$.

The time-dependence of the transverse magnetisation
$\langle\sigma_r^z(t)\rangle$ (which is in fact a piece of the
energy density) has been studied in Ref.~\cite{bm1}. For case of a
quench from $h_0>1$ to $h=1$ the results agree precisely with our
general results of a constant asymptotic value approached by an
oscillatory $t^{-3/2}$ power law, which we have argued above is
due to lattice effects. In Ref.~\cite{sps-04} the time averaged
behavior of the spin-spin correlation function was calculated for
quenches from $h_0=0$ and infinity. For quenches to the critical
theory with $h=1$, the behavior was found to be purely
exponential, in agreement with (\ref{greaterthan}).
In Ref.~\cite{ir-00} the behavior
of the correlation function in a finite system was studied
numerically. While in this case there are complications from both
the finite size and lattice effects, the main features are once
again consistent with our general theory.

We conclude with a discussion of the expected behavior in higher
dimensions $d>1$. For quenches from one disordered state to
another outside the critical region the gaussian model results
(\ref{2ptgaussian}) (with $\int dk\to\int d^dk$) should apply. For
$t<r/2v_m$ the 2-point function behaves as at $t=0$. At large
times it decays exponentially with a scale set by the correlation
length of $H$ (although its Fourier transform differs in detail
from the behavior in the ground state of $H$). The approach to
this for $t>r/2v_m$ is however more complicated, since the wave
fronts are no longer planar. A similar analysis may be applied to
the dynamics of the Goldstone modes, or spin waves, in the quench
to an ordered state. Above the upper critical dimension ($d=3$ if
$z=1$) mean field theory should also apply to quenches to the
critical point. In $d=3$, this gives $\langle\phi({\bf
r},t)\phi(0,t)\rangle\sim r^{-1}e^{-\pi(t-r/2v)/\tau_0}$ for
$t<r/2v$, and saturates at later times to $t$-independent value. 
Similarly we find,
analytically continuing the results in Ref.~\cite{ked-93} for a
slab geometry, that for a critical quench from a disordered state,
the energy density parameter $\langle\phi^2\rangle$ decays
exponentially. In principle it should also be possible to use the
results in \cite{kd-99} for a slab geometry in $d=3-\epsilon$
dimensions, but this is very cumbersome. The calculation is
simpler at large $N$ \cite{ked-93} and once again shows
exponential decay. However, a counter-example to this general rule
appears to be given by the behavior of the magnetization in a
critical quench from an ordered state. The mean-field profile in a
slab geometry, given in \cite{k-97}, involves elliptic functions,
which, when continued to real time, give non-decaying oscillations
with period $\sim\tau_0$. We expect this result
to be modified by the inclusion of fluctuations.

\acknowledgments This work was supported in part by EPSRC Grant
GR/R83712/01
and by the Stichting voor Fundamenteel Onderzoek der Materie (FOM).


\begin{thebibliography}{99}

\bibitem{bm1}
E. Barouch and B. McCoy,  Phys. Rev. A {\bf 2}, 1075 (1970);
{\bf 3}, 786 (1971); {\bf 3}, 2137 (1971).

\bibitem{ir-00}
F. Igloi and H. Rieger, Phys. Rev. Lett. {\bf 85}, 3233 (2000).
%[cond-mat/0003193].

\bibitem{uc}
M.~Greiner et al., Nature (London) {\bf 415}, 39 (2002);
C.~Orzel et al., Science {\bf 291}, 2386 (2001).

\bibitem{sps-04}
K. Sengupta et al.,  Phys. Rev. A {\bf 69},
053616 (2004).
%[cond-mat/0311355].

\bibitem{cl-05}
R. W. Cherng and L. S. Levitov, cond-mat/0512689.

\bibitem{dd}
H. W. Diehl, in {\sl Phase Transitions and
Critical Phenomena}, v.~10, ed C. Domb and J. L. Lebowitz 
(Academic, London, 1986); Int. J. Mod. Phys. B 11, 3503 (1997).

\bibitem{cardy-84}
J. L. Cardy, Nucl. Phys. B {\bf 240}, 514 (1984).

\bibitem{cardy-05} J. L. Cardy, Boundary Conformal Field Theory,
in {\sl Encyclopedia of Mathematical Physics}, ed J.-P. Francoise, G. Naber, 
and S. Tsun Tsou, (Elsevier, Amsterdam, 2006).


\bibitem{cc-05} P. Calabrese and J. Cardy,
J. Stat. Mech. P04010 (2005) [cond-mat/0503393].

\bibitem{cc-06} P. Calabrese and J. Cardy, in preparation.

\bibitem{cardybn-86} H. Bl\"ote et al.,
Phys. Rev. Lett. {\bf 56}, 742 (1986).

\bibitem{ked-93}
M. Krech et al., Phys. Rev. E {\bf 52},
1345, (1995).

\bibitem{kd-99}
R. Klimpel and S. Dietrich, Phys. Rev. B {\bf 60}, 16977 (1999).

\bibitem{k-97} M. Krech,
 Phys. Rev. E {\bf 56} 1642 (1997), and references therein.

\end{thebibliography}
\end{document}